\documentclass[11pt]{article}

\usepackage[margin=1.0in]{geometry}

\usepackage{graphicx}
\usepackage{dsfont}
\usepackage{amsmath,amsthm}
\usepackage{amssymb}
\usepackage{amsfonts}       
\usepackage{mathtools}
\usepackage{enumitem}
\usepackage{url}            
\usepackage{booktabs}       
\usepackage{nicefrac}       
\usepackage{microtype}      
\usepackage{natbib}
\usepackage{doi}
\usepackage{tikz}
\usepackage{tikz-cd}
\usepackage{appendix}
\usepackage{comment}
\usetikzlibrary{arrows.meta,positioning}
\usepackage{titletoc}
\usepackage{xcolor}
\usepackage{hyperref}
\hypersetup{
    colorlinks=true,
    linkcolor=blue,
    citecolor=blue,
    urlcolor=blue,
    filecolor=blue
}
\definecolor{covernavy}  {RGB}{  0,  32,  96}
\definecolor{inputbg} {RGB}{248, 249, 252}
\usepackage[capitalize,nameinlink]{cleveref}
\usepackage{pdflscape}
\usepackage{array}
\usepackage{multirow}
\usepackage{ragged2e}

\theoremstyle{plain}

\theoremstyle{definition}

\theoremstyle{remark}

\crefname{assumption}{Assumption}{Assumptions}
\Crefname{assumption}{Assumption}{Assumptions}
\crefname{condition}{Condition}{Conditions}
\Crefname{condition}{Condition}{Conditions}
\crefname{remark}{Remark}{Remarks}
\Crefname{remark}{Remark}{Remarks}

\DeclareMathOperator*{\argmin}{arg\,min}
\newcommand{\E}{\mathbb{E}}        
\newcommand{\PP}{\mathbb{P}}       
\newcommand{\one}{\mathds{1}}      

\renewcommand{\d}{\mathrm{d}}
\newcommand{\ind}{\perp\!\!\!\!\perp}

\begin{document}

\title{Marginal Structural Models for Electricity Demand under
Treatment-Confounder Feedback: A Continuous-Treatment Outcome-Adaptive and
Fused LASSO Approach}

\author{
\textbf{Shalini Jayanetti}\\
Department of Statistics,\\
University of Manitoba\\
\href{mailto:jayanesk@myumanitoba.ca}{jayanesk@myumanitoba.ca}
\and
\textbf{Sumeet Kalia}\\
Department of Statistics,\\
University of Manitoba\\
\href{mailto:Sumeet.Kalia@umanitoba.ca}{Sumeet.Kalia@umanitoba.ca}
}

\renewcommand{\thefootnote}{\ensuremath{\dagger}}
\maketitle
\renewcommand{\thefootnote}{\arabic{footnote}}
\setcounter{footnote}{0}

\begin{abstract}
Temperature is the most important meteorological driver of electricity demand, but
its causal effect has not been estimated under the time-varying confounding that
characterizes weather processes. Current temperature is associated with recent
weather conditions, including precipitation, snow cover, cloud cover, and air
density, that also influence electricity demand, and past temperature in turn shapes
future weather, producing treatment-confounder feedback under which standard
regression adjustment is biased, because conditioning on a time-varying confounder
that is also a mediator of past treatment removes part of the causal effect. We
estimate the causal effect of temperature on daily Ontario electricity demand from
2018 to 2019 using a marginal structural model (MSM) with inverse-probability
weighting (IPW), formulated for a single observed time series rather than a panel of
independent subjects, and we extend the longitudinal outcome-adaptive LASSO (LOAL)
and adaptive fused LASSO, previously developed for binary
treatments, to a continuous treatment using density-ratio weights and a
weighted-covariance balance criterion. In a Monte Carlo study calibrated to Ontario
weather and demand across effect sizes $\beta_1 \in \{0,2,4,8\}$, unadjusted
regression is biased toward the null with coverage of $0.12$ to $0.15$, whereas the
stabilized outcome-adaptive estimators are nearly unbiased (relative bias $-1.8\%$ at
$\beta_1 = 8$) with coverage near $0.92$, and the cumulative three-day estimators are
approximately unbiased but less efficient. In the Ontario data all estimators
identify a positive and highly significant quadratic temperature effect
($p < 0.001$); the unadjusted estimate is $9.34$ MW per squared degree Celsius, the
stabilized and outcome-adaptive single-lag estimators give $9.8$ to $10.1$, and the
cumulative estimators give smaller values ($5.2$ to $8.1$) that coincide with a sharp
fall in effective sample size and near-unit air-density collinearity ($r = -0.985$).
The temperature effect on demand is therefore large and robust to single-lag
confounding adjustment, while the cumulative estimates are compromised by a
structural positivity limitation that outcome-adaptive selection cannot remove.

\vspace{0.2in}

\end{abstract}


\newpage

\section{Introduction}

Reliable operation and planning of electricity systems rests on quantifying how
demand would respond to changes in its physical drivers, of which temperature is
the most important: heating and cooling load across the residential, commercial,
and industrial sectors produce a U-shaped response in which both cold and hot days
raise consumption \citep{valor2001daily, pardo2002temperature}. The nonlinearity of
this response is well documented, both in aggregate load models and in threshold
formulations that separate the heating and cooling regimes
\citep{bessec2008nonlinear, campbell2005weather}. Published analyses of the
temperature-demand relationship have, however, been almost exclusively predictive,
using dynamic regression with ARIMA errors and, more recently, machine learning to
forecast load \citep{taylor2003short, bowala2022superiority, bowala2024neural}.
These methods estimate association rather than the causal effect of a change in the
temperature regime, yet it is the latter that is relevant to scenario planning and
to anticipating demand under a changing climate, since a forecasting model
describes what demand will be under observed conditions whereas a causal model
describes how demand would change were those conditions set to specified values
\citep{hernan2020causal}.

Estimating the causal effect is complicated by the temporal structure of weather,
which induces time-varying confounding with feedback. Temperature on a given day is
associated with recent weather conditions, namely lagged precipitation, snowfall,
snow mass, cloud cover, and air density, and these variables also influence demand
directly through their effect on heating and cooling load, so that recent weather
confounds the temperature-demand relationship and its omission attributes to
temperature the demand variation that is in fact driven by correlated weather
\citep{wilks2011statistical}. Because past temperature affects subsequent weather,
such as snow mass and air density, which in turn affect both future temperature and
future demand, a lagged weather variable is simultaneously a confounder of the
current temperature effect and a mediator of the past temperature effect
\citep{robins1999association, vanderweele2009concerning}. Conditioning on such a
variable in an ordinary regression blocks the indirect pathway from past
temperature to current demand and biases the total effect, regardless of how
carefully the regression is specified. Marginal structural models (MSMs) estimated
by inverse-probability weighting (IPW) were developed for this situation:
rather than conditioning on the time-varying confounders, IPW re-weights the
observations to construct a pseudo-population in which treatment is independent of
the measured confounders while the causal effect is preserved
\citep{robins2000marginal, cole2008constructing}. The importance of the temporal
lag structure has a parallel in environmental epidemiology, where distributed-lag
nonlinear models are used to represent exposure-lag-response associations for
temperature and mortality \citep{gasparrini2010distributed}.

\subsection{Knowledge gap}
\label{sec:gap}

Two features of the temperature-demand problem fall outside the setting for which
MSMs and their software were developed, and neither has been addressed in the
energy-demand literature. The treatment, daily temperature, is continuous, so the
weights are ratios of conditional densities rather than inverse treatment
probabilities, and their construction and diagnostics differ accordingly
\citep{robins2000marginal, naimi2014constructing, hirano2004propensity,
huling2024independence}. More fundamentally, the available data form a single long
series for one unit, the Province of Ontario, rather than a sample of independent
subjects, whereas MSM theory and software presume many independent replicates.
Identifying and estimating a causal effect from one realized path requires that the
estimand be defined on the potential-outcome path of that unit and that inference
rely on stationarity and weak temporal dependence rather than on independent
sampling \citep{bojinov2019time, blackwell2018tscs}. A further obstacle is the
dimensionality of the treatment model, which must condition on the full lagged
weather history: with five weather variables at three lags the treatment model
contains fifteen strongly autocorrelated regressors, and the resulting
near-collinearity destabilizes the estimated weights and aggravates near-positivity violations
\citep{belsley1980regression, spreafico2024positivity}. Outcome-adaptive selection
addresses this last problem for point treatments by retaining confounders and
outcome predictors while discarding variables associated only with treatment
\citep{shortreed2017outcome}, and the outcome-adaptive LASSO has been extended to
longitudinal binary treatments through a longitudinal outcome-adaptive LASSO (LOAL)
followed by an adaptive fused LASSO that pools coefficients across time
\citep{schnitzer2026adaptive}. These procedures have not been formulated for
continuous treatments, and their behavior is known to degrade when candidate
covariates are strongly collinear \citep{balde2023reader}, as they are here.

\subsection{Objectives}
\label{sec:objectives}

This paper estimates the causal effect of temperature on daily Ontario electricity
demand and, in doing so, develops the methodology required to do so from a single
time series with a continuous, high-dimensional treatment model. We formulate the
MSM for the temperature-demand effect as a problem of causal inference in one
observed series, stating the estimand on the potential-outcome path of the unit
together with the identification and estimation conditions that this framing
requires, and we conduct inference with heteroskedasticity- and
autocorrelation-consistent (HAC) variance estimation in place of the
independence-based sandwich estimator \citep{newey1987simple}. We extend the LOAL
and adaptive fused LASSO of \citet{schnitzer2026adaptive} from binary to continuous
treatment by replacing the logistic treatment model and mean-difference balance
criterion with a Gaussian density-ratio weight and a weighted-covariance balance
criterion, and by defining the fusion graph across the lag-specific treatment
models. The estimators are evaluated in a Monte Carlo study calibrated to Ontario
weather and demand, in which treatment assignment depends on the confounders and
past treatment feeds back into future confounders, and the target parameter is
defined by g-computation. Applying the estimators to the Ontario series then yields
the first causal, as opposed to predictive, analysis of the temperature-demand
relationship, and quantifies a structural positivity limitation that arises because
air density is nearly a deterministic function of temperature.

\subsection{Outline}
\label{sec:outline}

Section~\ref{sec:notation} fixes notation, defines the causal estimand for a single
time series, and states the identification and estimation conditions.
Section~\ref{sec:methods} develops the continuous-treatment IPW estimator, the MSM,
the continuous LOAL and adaptive fused LASSO, and the inference procedure.
Section~\ref{sec:sim} describes the simulation design and reports the performance of
the thirteen estimators. Section~\ref{sec:application} presents the Ontario data,
the balance and weight diagnostics, and the estimated temperature effect.
Section~\ref{sec:discussion} discusses the findings, their connection to existing
work, and the limitations imposed by overlap. Derivations, the data-generating
process, and the causal diagram for the application are collected in the appendix.

\section{Notation and Causal Framework}
\label{sec:notation}

\subsection{Notation}
\label{sec:notation-sub}

We index days by $t = 1,\dots,T$, with $T = 730$ daily observations over the period
2018 to 2019. A subscript $t$ denotes a variable's value on day $t$ and $t-k$ its
value $k$ days earlier, and an overbar denotes history, so that
$\bar A_t = (A_1,\dots,A_t)$ and $\bar{\boldsymbol L}_t = (\boldsymbol L_1,\dots,
\boldsymbol L_t)$. The treatment $A_t \in \mathcal A \subseteq \mathbb R$ is the
daily mean temperature (\textdegree C), and $\bar a = (a_1,\dots,a_t)$ denotes a
candidate temperature regime. The outcome $Y_t$ is daily mean electricity demand
(MW). The vector $\boldsymbol L_t$ collects the five time-varying weather
confounders on day $t$, namely precipitation (mm), air density
(kg/m\textsuperscript{3}), snowfall (binary), snow mass (binary), and cloud cover.
Two exogenous predictors that are not affected by treatment are retained to improve
precision: the day-type indicator $D_t$ (weekday versus weekend or statutory
holiday) and the seasonal term $S_t = \cos\!\big(2\pi (d_t - 172)/365.25\big)$,
where $d_t$ is the day of year. The observed data on day $t$ are therefore
$\mathcal{O}_t = \{A_t, \boldsymbol L_t, Y_t, D_t, S_t\}$, with observed history
$\bar{\mathcal{O}}_t = (\mathcal{O}_1,\dots,\mathcal{O}_t)$, and the analysis is
based on the single realized path $\bar{\mathcal{O}}_T$.

In the potential-outcome time-series framework, $Y_t(\bar a)$ denotes the demand
that would be observed on day $t$ had the temperature regime been set to $\bar a$
\citep{bojinov2019time, robins1986new}. We write $f(a_t \mid \cdot)$ for the
conditional density of the continuous treatment and $\ind$ for conditional
independence. The intercept and slopes $\beta_0,\dots,\beta_3$ of the MSM introduced
below are marginal structural parameters, that is, population-averaged causal
effects in the IPW pseudo-population, and are not conditional regression
coefficients.

\subsection{Estimand}
\label{sec:estimand}

Because building thermal mass makes demand respond to recent rather than
instantaneous temperature, we index the exposure at the previous day's temperature
$A_{t-1}$, which is the lag used throughout, and absorb longer-range dynamics
through the confounder history; this single-lag choice mirrors the operative lag in
exposure-lag-response models for temperature \citep{gasparrini2010distributed}.
Writing the exposure generically as $a$, the marginal structural model for mean
demand is
\begin{equation}
\label{eq:msm}
\E\big[\,Y_t(a)\mid D_t, S_t\,\big]
   \;=\; \beta_0 \;+\; \beta_1\,(a - c)^2 \;+\; \beta_2 D_t \;+\; \beta_3 S_t ,
\end{equation}
where $c$ is a reference thermal-comfort temperature and $(a-c)^2$ encodes the
symmetric U-shaped heating and cooling load. The target causal parameter is the
curvature $\beta_1$, the population-averaged change in demand per unit squared
deviation of temperature from comfort. Conditioning on $D_t$ and $S_t$, which are
not descendants of treatment, does not require weighting and serves only to reduce
residual variance, while centering at $c$ in place of the raw quadratic $a^2$
locates the response minimum at the comfort point and reduces collinearity between
the quadratic term and the seasonal covariate; we set $c$ to the demand-minimizing
temperature estimated from the data.

The primary analysis targets the single-lag effect in \eqref{eq:msm}, whose weights
depend on the propensity for $A_{t-1}$. As an extension we consider the joint
three-day regime $\bar a = (a_{t-1},a_{t-2},a_{t-3})$, whose cumulative weights
intervene on the full three-day temperature history and therefore target a
different, joint estimand that is more sensitive to positivity violations. The two
regimes are kept distinct because cumulative weights are coherent only with an
outcome model that represents the joint effect they intervene on
\citep{hernan2001marginal, vanderweele2009concerning}.

\subsection{Identification and estimation conditions}
\label{sec:assumptions}

Identification of $\beta_1$ from a single observed series rests on the standard
assumptions of the longitudinal causal-inference literature, adapted here to a
continuous treatment \citep{robins2000marginal, hernan2020causal}. Consistency
requires that when the observed regime equals $\bar a$ the observed outcome equals
the corresponding potential outcome, $\bar A_t = \bar a \Rightarrow Y_t =
Y_t(\bar a)$. Sequential exchangeability, or no unmeasured confounding, requires
that at each day the treatment be independent of the future potential outcomes given
the treatment and confounder history, $Y_t(\bar a) \ind A_s \mid \bar A_{s-1},
\bar{\boldsymbol L}_s$ for all $s \le t$. Positivity, or overlap, requires that every
temperature value that occurs have positive conditional density given the history,
$f_{A_s \mid \bar A_{s-1}, \bar{\boldsymbol L}_s}(a_s) > 0$, bounded away from zero on
the region of interest, so that the density-ratio weights are well defined and have
finite variance. Positivity is the most restrictive of the three in this
application, because air density is, through the ideal-gas relation, a
near-deterministic function of temperature, so that conditioning on lagged air
density leaves little independent variation in the exposure; the practical
consequence, quantified in Section~\ref{sec:balance}, is extreme weights and
residual imbalance that no weighting scheme removes, which we treat as a structural
overlap limitation rather than as ordinary multicollinearity
\citep{petersen2012diagnosing}.

Because the data are a single realized path, identification and inference further
require that the joint process $\{(Y_t,A_t,\boldsymbol L_t)\}$ be covariance
stationary and $\alpha$-mixing, so that time averages converge to their expectations
and a central limit theorem holds for the HAC variance estimators used below
\citep{newey1987simple, bojinov2019time}. Stationarity is a property of the data
process rather than a causal assumption; under non-stationarity the pseudo-population
interpretation and the mean-one property of the stabilized weights need not hold and
estimates can be spurious \citep{box1976analysis, hamilton2020time}, so we assess it
empirically before estimation (Section~\ref{sec:application}).

\section{Methods}
\label{sec:methods}

\subsection{Inverse-probability weights for a continuous treatment}
\label{sec:ipw}

Under consistency, sequential exchangeability, positivity, and the stationarity and
weak-dependence conditions of Section~\ref{sec:assumptions}, the effect of the
exposure is recovered by weighting each day by the inverse of the conditional
treatment density, which removes the confounder-to-treatment dependence while
preserving the treatment effect (\cref{fig:DAG_EO}; the derivation is given in
Appendix~\ref{app:identification}). For the single-lag regime the unstabilized and
stabilized weights are
\begin{equation}
\label{eq:weights}
w_t \;=\; \frac{1}{f\!\left(A_{t-1}\mid \bar{\boldsymbol L}_{t-1}\right)},
\qquad
\mathrm{SW}_t \;=\;
\frac{f\!\left(A_{t-1}\mid \bar A_{t-2}\right)}
     {f\!\left(A_{t-1}\mid \bar A_{t-2}, \bar{\boldsymbol L}_{t-1}\right)} ,
\end{equation}
where the conditional density is modeled as Gaussian with mean from a linear
treatment model and variance from its residual mean square
\citep{robins2000marginal, naimi2014constructing}. Stabilization leaves the target
unchanged but reduces variance, since stabilized weights have mean approximately one
and lighter tails \citep{cole2008constructing, xu2010use}. For the joint three-day
regime the cumulative weights multiply the day-specific density ratios over lags
$t-1$, $t-2$, and $t-3$; because these weights intervene on a richer regime they are
more sensitive to near-positivity and are reported as an extension rather than as the
primary estimator. Extreme weights are trimmed at the 95th percentile, and we report
the effective sample size $(\sum_t w_t)^2/\sum_t w_t^2$ as a stability diagnostic
\citep{cole2008constructing}.

\begin{landscape}
\begin{figure}
    \centering
\scalebox{0.85}{\begin{tikzpicture}[
  > = {Latex[length=2mm, width=2mm]},
  shorten > = 1pt,
  auto,
  scale = 1.2,
  node distance = 4cm,
  semithick
]
\tikzstyle{state}=[
  circle,
  draw = black,
  thick,
  minimum size = 12mm,
  inner sep = 2pt,
  align = center
]
\node[font=\bfseries, anchor=west] at (-1.5, 4) {Observational ($\mathcal{O}$)};
\node[state] (L0) at (0, 3) {$\mathbf{L}_0$};
\node[state] (L1) at (4, 3) {$\mathbf{L}_1$};
\node[state] (L2) at (8, 3) {$\mathbf{L}_2$};
\node[state] (A0) at (1, 0) {$A_0$};
\node[state] (A1) at (5, 0) {$A_1$};
\node[state] (A2) at (9, 0) {$A_2$};
\node[state] (Y1) at (3, 0) {$Y_1$};
\node[state] (Y2) at (7, 0) {$Y_2$};
\node[state] (Y3) at (11, 0) {$Y_3$};
\draw[->, thick, gray] (L0) -- (L1);
\draw[->, thick, gray] (L1) -- (L2);
\draw[->, thick, red] (L0) -- (A0);
\draw[->, thick, red] (L1) -- (A1);
\draw[->, thick, red] (L2) -- (A2);
\draw[->, thick, gray] (L0) -- (Y1);
\draw[->, thick, gray] (L1) -- (Y2);
\draw[->, thick, gray] (L2) -- (Y3);
\draw[->, thick, red] (A0) to[bend left=20] (L1);
\draw[->, thick, red] (A1) to[bend left=20] (L2);
\draw[->, thick, gray] (A0) -- (Y1);
\draw[->, thick, gray] (A1) -- (Y2);
\draw[->, thick, gray] (A2) -- (Y3);
\draw[->, thick, gray] (Y1) to[bend right=30] (Y3);
\draw[->, thick, gray] (Y1) to[bend right=20] (Y2);
\draw[->, thick, gray] (Y2) to[bend right=30] (Y3);
\draw[->, thick, gray] (L0) -- (A1);
\draw[->, thick, gray] (L1) -- (A2);
\draw[->, thick, gray] (L0) -- (A2);
\end{tikzpicture}} \\
\scalebox{0.85}{\begin{tikzpicture}[
  > = {Latex[length=2mm, width=2mm]},
  shorten > = 1pt,
  auto,
  scale = 1.2,
  node distance = 4cm,
  semithick
]
\tikzstyle{state}=[
  circle,
  draw = black,
  thick,
  minimum size = 12mm,
  inner sep = 2pt,
  align = center
]
\node[font=\bfseries, anchor=west] at (-1.5, 4) {Experimental ($\mathcal{E}$)};
\node[state] (L0) at (0, 3) {$\mathbf{L}_0$};
\node[state] (L1) at (4, 3) {$\mathbf{L}_1$};
\node[state] (L2) at (8, 3) {$\mathbf{L}_2$};
\node[state] (A0) at (1, 0) {$A_0$};
\node[state] (A1) at (5, 0) {$A_1$};
\node[state] (A2) at (9, 0) {$A_2$};
\node[state] (Y1) at (3, 0) {$Y_1$};
\node[state] (Y2) at (7, 0) {$Y_2$};
\node[state] (Y3) at (11, 0) {$Y_3$};
\draw[->, thick, gray] (L0) -- (L1);
\draw[->, thick, gray] (L1) -- (L2);
\draw[->, thick, gray] (L0) -- (Y1);
\draw[->, thick, gray] (L1) -- (Y2);
\draw[->, thick, gray] (L2) -- (Y3);
\draw[->, thick, red] (A0) to[bend left=20] (L1);
\draw[->, thick, red] (A1) to[bend left=20] (L2);
\draw[->, thick, gray] (A0) -- (Y1);
\draw[->, thick, gray] (A1) -- (Y2);
\draw[->, thick, gray] (A2) -- (Y3);
\draw[->, thick, gray] (Y1) to[bend right=30] (Y3);
\draw[->, thick, gray] (Y1) to[bend right=20] (Y2);
\draw[->, thick, gray] (Y2) to[bend right=30] (Y3);

\end{tikzpicture}}
\caption{Directed acyclic graph for a time-varying treatment with
treatment-confounder feedback, under the observational regime ($\mathcal{O}$, top)
and the hypothetical randomized regime ($\mathcal{E}$, bottom). Red arrows mark the
confounding paths $\boldsymbol L_t \to A_t$ and the feedback paths
$A_{t-1}\to\boldsymbol L_t$. Inverse-probability weighting constructs a
pseudo-population in which the $\boldsymbol L_t \to A_t$ arrows are removed, so that
the association between treatment and outcome recovers the causal effect. Three time
points are shown for illustration; the study uses $T=730$ autocorrelated daily
observations.}
\label{fig:DAG_EO}
\end{figure}
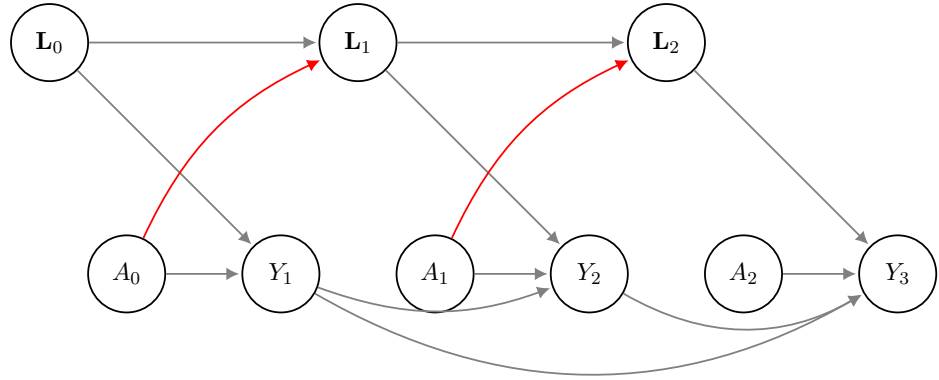
\end{landscape}

\subsection{Marginal structural model}
\label{sec:msm}

The MSM \eqref{eq:msm} is fitted by weighted least squares, regressing $Y_t$ on
$(A_{t-1}-c)^2$, $D_t$, and $S_t$ with weights $w_t$ or $\mathrm{SW}_t$. Because the
weights account for the time-varying confounders, the confounder history does not
enter the outcome regression directly, and $\beta_1$ is the marginal structural
curvature in the weighted pseudo-population. Since the data form a single dependent
series, standard errors are HAC (Newey-West) estimators \citep{newey1987simple},
supplemented by a moving-block bootstrap as a robustness check
\citep{morris2019using}.

\subsection{Longitudinal outcome-adaptive LASSO for a continuous treatment}
\label{sec:loal}

The propensity model conditions on fifteen strongly autocorrelated lagged weather
regressors, so including all of them without regard to their role inflates the
variance of the weights without reducing bias, and the outcome-adaptive LASSO is
known to deteriorate under exactly this kind of collinearity
\citep{shortreed2017outcome, balde2023reader}. We therefore adapt the longitudinal
outcome-adaptive LASSO of \citet{schnitzer2026adaptive} to select confounders for a
continuous treatment. A working outcome model first regresses $Y$ on treatment and
all candidate covariates to obtain outcome-adaptive coefficients
$\hat{\boldsymbol\beta}$, whose magnitudes measure each covariate's association with
the outcome and set the adaptive penalty weights $\hat\omega_j =
|\hat\beta_j|^{-\gamma}$ with $\gamma = 2.5$, a value within the range for which the
adaptive LASSO has the oracle property, so that covariates strongly associated
with the outcome are penalized least \citep{zou2006adaptive}. A pooled treatment
model over the lag blocks is then estimated by the adaptive LASSO,
\begin{equation}
\label{eq:loal}
\hat{\boldsymbol\alpha}(\lambda_n)
 = \argmin_{\boldsymbol\alpha}\Bigg[
   \sum_{\tau}\sum_{i} \ell\!\left(A_{\tau,i},\, \boldsymbol\alpha^{\!\top}
   \mathbf V_{\tau,i}\right)
   + \lambda_n \sum_{j\in\mathcal J} \hat\omega_j\,|\alpha_j|\Bigg],
\end{equation}
where $\mathcal J$ indexes the penalized confounder coefficients, excluding
intercepts and treatment terms. The penalty $\lambda_n$ is chosen not by
cross-validation but by a longitudinal balance criterion,
\begin{equation}
\label{eq:bc}
\mathrm{BC}(\hat{\boldsymbol\alpha}) =
\sum_{j\in\mathcal J} |\hat\beta_j|\,
\big|\mathrm{WBAL}_j(\hat{\boldsymbol\alpha})\big| ,
\end{equation}
minimized over a grid, where $\mathrm{WBAL}_j$ is a covariate-$j$ balance measure
under the candidate weights \citep{shortreed2017outcome, schnitzer2026adaptive}.

Three features distinguish this continuous-treatment version from its binary
predecessor. The logistic treatment model $\ell(\cdot)$ is replaced by a Gaussian
squared-error model, from which the conditional density needed for the weights is
obtained as a normal density with mean from the fitted model and variance from its
residual mean square \citep{naimi2014constructing}. Because a continuous treatment
admits uncountably many regimes, the outcome-adaptive coefficients
$\hat{\boldsymbol\beta}$ are obtained from a single working outcome model on the
observed treatment and confounder history rather than by enumerating regimes.
Balance for a continuous treatment cannot be measured by a mean difference between
treated and untreated groups, so $\mathrm{WBAL}_j$ is taken to be the absolute
weighted correlation between the treatment and covariate $j$, which is zero when the
weighted treatment-covariate association is removed \citep{austin2015moving}.

\subsection{Adaptive fused LASSO for a continuous treatment}
\label{sec:floal}

After selection, coefficients of the same covariate can be pooled across the
lag-specific treatment models by an adaptive fused LASSO, which reduces the effective
number of parameters and improves efficiency under sparsity
\citep{schnitzer2026adaptive}. The three density ratios in the cumulative weight come
from three treatment models, for $A_{t-1}$, $A_{t-2}$, and $A_{t-3}$, that share the
same confounder design, and under stationarity the effect of a given confounder on
temperature need not differ across these models. We therefore let the fusion graph
$\mathcal E$ connect the coefficient of each confounder across the three
lag-specific models, without fusing distinct confounders, and solve
\begin{equation}
\label{eq:floal}
\argmin_{\boldsymbol\alpha^{*}}\Bigg[
 \sum_{\tau}\sum_{i} \ell\!\left(A_{\tau,i},\,
 \boldsymbol\alpha^{*\top}\mathbf V_{\tau,i}\right)
 + \lambda_{1,n}\!\!\sum_{(k,k')\in\mathcal E}\!
   \frac{|\alpha^{*}_{k'} - \alpha^{*}_{k}|}
        {\,|\hat\alpha^{\mathrm{refit}}_{k'}-\hat\alpha^{\mathrm{refit}}_{k}|^{\gamma_1}}
\Bigg],
\end{equation}
with $\gamma_1 = 2.5$ and $\lambda_{1,n}$ chosen by BIC \citep{viallon2016robustness}.
When a confounder's three coefficients are fused they are represented by a single
lag-invariant parameter and the model is refitted without penalty, which is the
direct analogue, for the three treatment models entering the cumulative weight, of
pooling a covariate's coefficient across time points in \citet{schnitzer2026adaptive}.
No further sparsity penalty is imposed, since selection is completed in the LOAL step.

\subsection{Inference under selection}
\label{sec:inference}

For the fixed-propensity estimators the propensity model is pre-specified, so the HAC
standard errors are valid under the stationarity and weak-dependence conditions
\citep{newey1987simple}. For the LOAL and fused-LOAL estimators the propensity model
is selected within the analysis, and a plug-in sandwich estimator then understates
uncertainty by ignoring the selection step \citep{schnitzer2026adaptive}; their
standard errors and confidence intervals are therefore obtained by an
$m$-out-of-$n$ moving-block bootstrap that re-runs the entire
selection-and-weighting pipeline within each resample, which respects both the
temporal dependence and the data-adaptive model choice. The variance estimator used
is stated in each table so that the fixed and adaptive families are not conflated.

\section{Simulation Study}
\label{sec:sim}

\subsection{Design}
\label{sec:sim-design}

We evaluate the estimators by Monte Carlo simulation with $B = 1{,}000$ replications
calibrated to Ontario weather and demand. Each replication generates $N = 730$ daily
observations, of which the first three are removed for lag construction, leaving
$N = 727$ for estimation. The data-generating process, specified in full in
Appendix~\ref{app:dgp} and Table~\ref{tab:params}, is constructed so that the
confounding the method is meant to remove is present: each weather confounder
follows an AR(1) process with a seasonal mean and feedback from the previous day's
temperature, $A_{t-1}\to\boldsymbol L_t$; temperature depends on the contemporaneous
confounders together with the seasonal term and noise, $\boldsymbol L_t \to A_t$; and
demand depends on the previous day's temperature through $(A_{t-1}-c)^2$, on the
lagged confounders $\boldsymbol L_{t-1}$, and on its own recent lags. Because
$\boldsymbol L_{t-1}$ affects both $A_{t-1}$ and $Y_t$ it confounds the
$A_{t-1}\to Y_t$ effect, and because $A_{t-1}$ feeds back into $\boldsymbol L_t$ the
confounders are also mediators of earlier temperature, reproducing the structure of
\cref{fig:DAG_EO}. The treatment noise variance is set large enough that overlap
holds by construction, so the simulation isolates estimator performance under valid
positivity; the near-positivity encountered in the application arises from a feature
of the real data, the near-deterministic dependence of air density on temperature,
that the simulation does not impose. The true value of $\beta_1$ is defined by
g-computation, computing $\E[Y_t(a)]$ over a grid of $a$ by intervening on
temperature (severing $\boldsymbol L_t\to A_t$) and averaging the simulated potential
outcomes, then recovering $\beta_1$ from \eqref{eq:msm}
\citep{robins1986new, morris2019using}. Under the additive data-generating process
the marginal and structural curvatures coincide, which we verify numerically, and the effect sizes examined are $\beta_1 \in \{0,2,4,8\}$.

We compare thirteen estimators in three families (Table~\ref{tab:estimators}): five
fixed-propensity estimators (No IPW, IPW, sIPW, cumulative IPW, and cumulative sIPW);
four LOAL variants that apply these weighting schemes after outcome-adaptive
selection; and four fused-LOAL variants that add the fusion step. All use the outcome
model \eqref{eq:msm}. For each estimator we report the mean estimate, the relative
bias as a percentage (omitted when $\beta_1 = 0$), the Monte Carlo standard deviation
(empirical standard error), the mean squared error, and the coverage of the nominal
95\% confidence interval; these are the standard measures for simulation studies of
estimator performance \citep{morris2019using}. Monte Carlo standard errors, computed
for each measure, are approximately $0.03$ for the bias and $0.01$ for the coverage
at $B = 1{,}000$, so the differences discussed below exceed simulation noise. For
tractability, coverage in the simulation uses the plug-in HAC standard error for all
thirteen estimators, whereas the moving-block bootstrap of
Section~\ref{sec:inference}, which additionally accounts for the selection step, is
reserved for the adaptive estimators in the application, where a single dataset makes
it feasible.

\begin{table}[ht]
\centering
\caption{The thirteen estimators, in three families. Weights are trimmed at the
95th percentile and density-ratio weights use a Gaussian treatment model. HAC
standard errors are used for the fixed-propensity family and moving-block bootstrap
for the LOAL and fused-LOAL families.}
\label{tab:estimators}
\renewcommand{\arraystretch}{1.2}
\small
\begin{tabular}{lll}
\toprule
\textbf{Family} & \textbf{Estimator} & \textbf{Weight construction} \\
\midrule
\multirow{5}{*}{Fixed propensity}
 & No IPW          & none (unadjusted regression) \\
 & IPW             & single-lag unstabilized density ratio \\
 & sIPW            & single-lag stabilized density ratio \\
 & Cum.\ IPW       & product of unstabilized ratios over lags 1--3 \\
 & Cum.\ sIPW      & product of stabilized ratios over lags 1--3 \\
\midrule
\multirow{4}{*}{LOAL}
 & LOAL-IPW        & IPW after outcome-adaptive selection \\
 & LOAL-sIPW       & sIPW after outcome-adaptive selection \\
 & LOAL-Cum.\ IPW  & cumulative IPW after selection \\
 & LOAL-Cum.\ sIPW & cumulative sIPW after selection \\
\midrule
\multirow{4}{*}{Fused LOAL}
 & FLOAL-IPW       & IPW after selection and fusion \\
 & FLOAL-sIPW      & sIPW after selection and fusion \\
 & FLOAL-Cum.\ IPW & cumulative IPW after selection and fusion \\
 & FLOAL-Cum.\ sIPW& cumulative sIPW after selection and fusion \\
\bottomrule
\end{tabular}
\end{table}

\subsection{Results}
\label{sec:sim-results}

Table~\ref{tab:sim_all} reports the performance of the thirteen estimators across
the four effect sizes. Unadjusted regression is severely biased toward the null:
because the confounding induces an approximately constant absolute bias of about
$-2.65$, its relative bias shrinks as $\beta_1$ grows, from $-131\%$ at $\beta_1 = 2$
to $-66\%$ at $\beta_1 = 4$ and $-33\%$ at $\beta_1 = 8$, while its coverage remains
far below nominal, between $0.12$ and $0.15$, throughout, which confirms that the
confounding is strong and that adjustment is necessary. Single-lag weighting removes
most of this bias and raises coverage, and the stabilized estimator sIPW improves on
the unstabilized IPW, consistent with the lighter tails of stabilized weights,
attaining relative bias of $-46\%$, $-23\%$, and $-12\%$ at $\beta_1 = 2$, $4$, and
$8$ with coverage near $0.90$. The outcome-adaptive stabilized estimators LOAL-sIPW
and FLOAL-sIPW perform best, reducing the relative bias to $-8.8\%$, $-3.7\%$, and
$-1.8\%$ across the same scenarios, with coverage of $0.89$ to $0.92$ and the
smallest mean squared error among all estimators; LOAL and FLOAL coincide for the
single-lag schemes because fusion acts only across the three lag models of the
cumulative weight. The cumulative estimators are approximately unbiased on average
but less efficient, with Monte Carlo standard deviations between $1.0$ and
$1.9$ against roughly $0.8$ for the single-lag schemes, and their coverage is lower
and more variable, particularly for the unstabilized cumulative variants; this loss
of efficiency is the price of intervening on the three-day regime and foreshadows the
overlap problem seen in the application. Overall, stabilization improves coverage over
the unstabilized weights and outcome-adaptive selection reduces bias further at no
cost in variance, so the stabilized adaptive estimators are preferred.

\begin{landscape}
\begin{table}[ht]
\centering
\caption{Simulation results across four effect-size scenarios. Entries are the mean estimate, relative bias (\%; omitted when $\beta_1=0$), Monte Carlo standard deviation (MCSD), mean squared error (MSE), and coverage of the nominal 95\% interval, from $B=1{,}000$ replications. Monte Carlo standard errors are approximately $0.03$ for the bias and $0.01$ for the coverage.}
\label{tab:sim_all}
\renewcommand{\arraystretch}{1.1}
\setlength{\tabcolsep}{4pt}
\scriptsize

\vspace{0.1in}

\begin{minipage}[t]{0.47\textwidth}
\centering
\textbf{Scenario $\beta_1=0$}

\begin{tabular}{lccccc}
\toprule
Estimator & Mean & RelBias\% & MCSD & MSE & Coverage\\
\midrule
No IPW           & -2.662 & --- & 0.858 & 7.822 & 0.145 \\
IPW              & -1.013 & --- & 0.836 & 1.724 & 0.824 \\
sIPW             & -0.973 & --- & 0.832 & 1.640 & 0.891 \\
Cum IPW          & -0.967 & --- & 1.108 & 2.162 & 0.847 \\
Cum sIPW         & -1.100 & --- & 1.220 & 2.699 & 0.858 \\
LOAL IPW         & -0.883 & --- & 1.043 & 1.869 & 0.806 \\
LOAL sIPW        & -0.180 & --- & 1.103 & 1.250 & 0.916 \\
LOAL Cum IPW     & -1.285 & --- & 1.335 & 3.433 & 0.769 \\
LOAL Cum sIPW    & -0.409 & --- & 1.711 & 3.096 & 0.896 \\
FLOAL IPW        & -0.883 & --- & 1.043 & 1.869 & 0.806 \\
FLOAL sIPW       & -0.180 & --- & 1.103 & 1.250 & 0.916 \\
FLOAL Cum IPW    & -0.935 & --- & 1.340 & 2.670 & 0.841 \\
FLOAL Cum sIPW   & 0.356 & --- & 1.729 & 3.117 & 0.922 \\
\bottomrule
\end{tabular}
\end{minipage}
\hspace{0.15\textwidth}
\begin{minipage}[t]{0.47\textwidth}
\centering
\textbf{Scenario $\beta_1=2$}

\begin{tabular}{lccccc}
\toprule
Estimator & Mean & RelBias\% & MCSD & MSE & Coverage\\
\midrule
No IPW           & -0.625 & -131.269 & 0.873 & 7.654 & 0.149 \\
IPW              & 1.038 & -48.077 & 0.852 & 1.651 & 0.826 \\
sIPW             & 1.088 & -45.587 & 0.852 & 1.557 & 0.898 \\
Cum IPW          & 1.015 & -49.255 & 1.095 & 2.170 & 0.818 \\
Cum sIPW         & 0.941 & -52.929 & 1.285 & 2.773 & 0.859 \\
LOAL IPW         & 1.132 & -43.400 & 1.059 & 1.876 & 0.811 \\
LOAL sIPW        & 1.825 & -8.748 & 1.146 & 1.345 & 0.891 \\
LOAL Cum IPW     & 0.619 & -69.038 & 1.352 & 3.735 & 0.751 \\
LOAL Cum sIPW    & 1.528 & -23.590 & 1.823 & 3.546 & 0.862 \\
FLOAL IPW        & 1.132 & -43.400 & 1.059 & 1.876 & 0.811 \\
FLOAL sIPW       & 1.825 & -8.748 & 1.146 & 1.345 & 0.891 \\
FLOAL Cum IPW    & 0.966 & -51.703 & 1.367 & 2.939 & 0.813 \\
FLOAL Cum sIPW   & 2.294 & 14.682 & 1.860 & 3.545 & 0.899 \\
\bottomrule
\end{tabular}
\end{minipage}

\vspace{0.8cm}

\begin{minipage}[t]{0.47\textwidth}
\centering
\textbf{Scenario $\beta_1=4$}

\begin{tabular}{lccccc}
\toprule
Estimator & Mean & RelBias\% & MCSD & MSE & Coverage\\
\midrule
No IPW           & 1.355 & -66.115 & 0.867 & 7.746 & 0.152 \\
IPW              & 3.029 & -24.270 & 0.819 & 1.613 & 0.854 \\
sIPW             & 3.080 & -23.010 & 0.845 & 1.561 & 0.893 \\
Cum IPW          & 2.989 & -25.279 & 1.039 & 2.103 & 0.856 \\
Cum sIPW         & 2.906 & -27.356 & 1.219 & 2.684 & 0.857 \\
LOAL IPW         & 3.149 & -21.277 & 1.057 & 1.841 & 0.822 \\
LOAL sIPW        & 3.854 & -3.661 & 1.119 & 1.274 & 0.921 \\
LOAL Cum IPW     & 2.597 & -35.078 & 1.347 & 3.782 & 0.757 \\
LOAL Cum sIPW    & 3.518 & -12.059 & 1.778 & 3.393 & 0.875 \\
FLOAL IPW        & 3.149 & -21.277 & 1.057 & 1.841 & 0.822 \\
FLOAL sIPW       & 3.854 & -3.661 & 1.119 & 1.274 & 0.921 \\
FLOAL Cum IPW    & 2.959 & -26.029 & 1.352 & 2.911 & 0.834 \\
FLOAL Cum sIPW   & 4.321 & 8.029 & 1.832 & 3.458 & 0.907 \\
\bottomrule
\end{tabular}
\end{minipage}
\hspace{0.15\textwidth}
\begin{minipage}[t]{0.47\textwidth}
\centering
\textbf{Scenario $\beta_1=8$}

\begin{tabular}{lccccc}
\toprule
Estimator & Mean & RelBias\% & MCSD & MSE & Coverage\\
\midrule
No IPW           & 5.336 & -33.298 & 0.850 & 7.819 & 0.120 \\
IPW              & 7.008 & -12.398 & 0.814 & 1.646 & 0.838 \\
sIPW             & 7.058 & -11.770 & 0.812 & 1.547 & 0.903 \\
Cum IPW          & 6.978 & -12.779 & 1.049 & 2.146 & 0.842 \\
Cum sIPW         & 6.885 & -13.935 & 1.200 & 2.683 & 0.873 \\
LOAL IPW         & 7.161 & -10.483 & 1.058 & 1.822 & 0.824 \\
LOAL sIPW        & 7.855 & -1.808 & 1.114 & 1.262 & 0.911 \\
LOAL Cum IPW     & 6.594 & -17.573 & 1.295 & 3.654 & 0.770 \\
LOAL Cum sIPW    & 7.557 & -5.539 & 1.686 & 3.039 & 0.902 \\
FLOAL IPW        & 7.161 & -10.483 & 1.058 & 1.822 & 0.824 \\
FLOAL sIPW       & 7.855 & -1.808 & 1.114 & 1.262 & 0.911 \\
FLOAL Cum IPW    & 6.910 & -13.623 & 1.324 & 2.940 & 0.835 \\
FLOAL Cum sIPW   & 8.272 & 3.400 & 1.766 & 3.192 & 0.922 \\
\bottomrule
\end{tabular}
\end{minipage}

\end{table}
\end{landscape}

\section{Application}
\label{sec:application}

\subsection{Data}
\label{sec:app-data}

\begin{landscape}
\begin{figure}[htbp]
\centering
\includegraphics[scale=0.6]{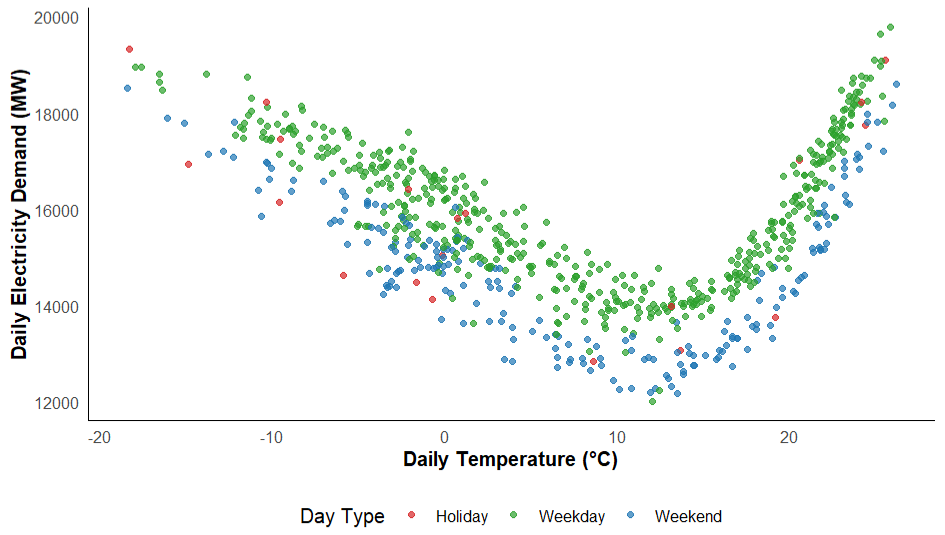}
\caption{Daily electricity demand versus daily mean temperature, Ontario, January
2018 to December 2019. The U-shaped envelope reflects heating load at low
temperatures and cooling load at high temperatures and motivates the centered
quadratic specification in \eqref{eq:msm}.}
\label{fig:tempDemand_Sactterplot}
\end{figure}
\end{landscape}

We analyze daily electricity demand and weather for Ontario, Canada, from January
2018 to December 2019 (\cref{fig:tempDemand_Sactterplot}). Demand was obtained from
the provincial system operator and the weather variables from a bias-corrected
reanalysis product \citep{staffell2016using}. Hourly demand and weather were
aggregated to daily means, snowfall and snow mass were dichotomized at zero, and
precipitation was log-transformed. The exposure is daily mean temperature and the
outcome is daily mean demand; the confounders are the five weather variables at lags
one to three, and the exogenous predictors are day type and the seasonal term
(Section~\ref{sec:estimand}). The reference temperature, estimated as the
demand-minimizing point of a quadratic fit, is $c = 8.1$\textdegree C. The causal
structure assumed for the application is shown in \cref{fig:DAG_1}.

Stationarity was assessed before estimation by deseasonalizing demand and
temperature and applying the augmented Dickey-Fuller test to the residual series,
which rejects the unit-root null for both demand ($p = 0.014$) and temperature
($p = 0.010$), supporting the weak-dependence condition of
Section~\ref{sec:assumptions} \citep{box1976analysis, hamilton2020time}.

\subsection{Balance and weight diagnostics}
\label{sec:balance}

The balance diagnostic reported in \cref{fig:Balance assessment} of the appendix
shows covariate balance as the absolute weighted correlation between the exposure $A_{t-1}$ and each lagged confounder across the thirteen weighting schemes, with covariates ordered by their unweighted correlation.
The air-density lags exhibit near-unit unweighted correlation with temperature
($r = -0.985$ between contemporaneous air density and temperature) and are not
brought below the conventional threshold $|r| < 0.1$ by any scheme, adaptive or not,
which is the empirical counterpart of the structural overlap limitation of
Section~\ref{sec:assumptions}: because air density is nearly a deterministic function
of temperature, no weighting can create the independent variation that overlap
requires \citep{petersen2012diagnosing}. Precipitation and cloud-cover lags show
moderate imbalance that all IPW schemes substantially reduce, and the binary snow
variables are well balanced, particularly under the cumulative schemes. The LOAL and
fused-LOAL schemes attain balance profiles comparable to their fixed-propensity
counterparts while retaining a sparser propensity model.

The effective sample size makes the overlap problem quantitative. Against the $727$
available days, the single-lag schemes retain a usable sample of $503$ (IPW), $443$
(sIPW), and $438$ (LOAL-sIPW), whereas the cumulative schemes collapse to between
$91$ and $121$, a loss driven by the near-collinearity of the lagged weather
variables and compounded in the cumulative weights, which multiply three treatment
models \citep{belsley1980regression, chatfield2019analysis}. The cumulative
estimators are therefore based on roughly one-sixth of the effective information of
the single-lag schemes, which bears directly on the interpretation of their point
estimates below.

\subsection{Causal effect estimates}
\label{sec:app-estimates}

Table~\ref{tab:msm_results} reports the estimated quadratic temperature effect
$\hat\beta_1$ for the thirteen schemes, with HAC standard errors for the
fixed-propensity family and moving-block bootstrap standard errors for the LOAL and
fused-LOAL families. All thirteen estimators identify a positive and highly
significant effect ($p < 0.001$ throughout), confirming the U-shaped
temperature-demand response after adjustment for weather confounding. The unadjusted
estimate is $\hat\beta_1 = 9.34$ MW per squared degree Celsius. The single-lag
adjusted estimators are close to this value, at $8.82$ (IPW), $10.08$ (sIPW), and
$9.80$ (LOAL-sIPW and FLOAL-sIPW), so that single-lag confounding adjustment changes
the estimate by at most about one unit; the slight increase under the stabilized and
adaptive estimators is in the direction predicted by the simulation, in which
unadjusted regression is biased toward the null. The cumulative schemes give smaller
estimates, at $7.67$ (cumulative IPW), $8.10$ (cumulative sIPW), and $5.23$ (LOAL and
FLOAL cumulative), but these coincide with the sharp drop in effective sample size
documented in Section~\ref{sec:balance} and with the near-unit air-density
collinearity, so the attenuation is attributable to the positivity limitation rather
than to a cleaner adjustment for confounding. Taken together with the simulation,
which identifies the stabilized outcome-adaptive single-lag estimators as least
biased, the analysis points to a temperature effect of approximately $9.8$ MW per
squared degree Celsius, obtained from a propensity model that outcome-adaptive
selection renders substantially sparser than the full fifteen-regressor model.

\begin{table}[htbp]
\centering
\caption{Estimated quadratic temperature effect on Ontario electricity demand (MW
per squared degree Celsius; 95th-percentile weight trimming). Fixed-propensity
estimators use HAC (Newey-West) standard errors; LOAL and fused-LOAL estimators use
moving-block bootstrap standard errors.}
\label{tab:msm_results}
\renewcommand{\arraystretch}{1.15}
\begin{tabular}{lrrrrrl}
\toprule
Estimator & Est. & SE & 95\% CI low & 95\% CI high & $z$ & Sig. \\
\midrule
\multicolumn{7}{l}{\textit{Fixed-propensity estimators (HAC SE)}} \\
No IPW           &  9.340 & 1.039 &  7.304 & 11.376 &  8.991 & *** \\
IPW              &  8.823 & 0.971 &  6.920 & 10.726 &  9.088 & *** \\
sIPW             & 10.083 & 0.894 &  8.332 & 11.835 & 11.285 & *** \\
Cumulative IPW   &  7.669 & 0.707 &  6.283 &  9.055 & 10.847 & *** \\
Cumulative sIPW  &  8.098 & 0.815 &  6.502 &  9.695 &  9.941 & *** \\
\midrule
\multicolumn{7}{l}{\textit{LOAL estimators (block-bootstrap SE)}}\\
LOAL-IPW         &  7.937 & 1.255 &  5.477 & 10.397 &  6.324 & *** \\
LOAL-sIPW        &  9.803 & 1.834 &  6.208 & 13.399 &  5.344 & *** \\
LOAL-Cum.\ IPW   &  5.231 & 1.153 &  2.971 &  7.490 &  4.537 & *** \\
LOAL-Cum.\ sIPW  &  5.972 & 1.398 &  3.232 &  8.713 &  4.272 & *** \\
\midrule
\multicolumn{7}{l}{\textit{Fused-LOAL estimators (block-bootstrap SE)}} \\
FLOAL-IPW        &  7.937 & 1.186 &  5.613 & 10.261 &  6.694 & *** \\
FLOAL-sIPW       &  9.803 & 1.864 &  6.149 & 13.458 &  5.258 & *** \\
FLOAL-Cum.\ IPW  &  5.231 & 1.204 &  2.871 &  7.591 &  4.344 & *** \\
FLOAL-Cum.\ sIPW &  5.972 & 1.418 &  3.193 &  8.752 &  4.211 & *** \\
\midrule
\multicolumn{7}{l}{\footnotesize $^{***}\,p<0.001$.} \\
\end{tabular}
\end{table}

\section{Discussion}
\label{sec:discussion}

We have estimated the causal effect of temperature on Ontario electricity demand
under treatment-confounder feedback, using a marginal structural model with
inverse-probability weighting adapted to two features of the problem that are
usually absent from epidemiological applications, a continuous treatment and a single
long time series. Ordinary regression is biased here because the lagged weather
variables are simultaneously confounders of the current temperature effect and
mediators of past temperature, and inverse-probability weighting resolves this by
re-weighting rather than conditioning; the simulation makes the consequence concrete,
since unadjusted regression is biased toward the null throughout, from two-thirds at
an effect size of four to a third at an effect size of eight, while the stabilized
outcome-adaptive estimators recover the target to within a few percent. To keep the high-dimensional, autocorrelated propensity model
tractable we extended the longitudinal outcome-adaptive LASSO and adaptive fused
LASSO of \citet{schnitzer2026adaptive} to a continuous treatment, using density-ratio
weights, a weighted-covariance balance criterion, and fusion across the lag-specific
treatment models, and this outcome-adaptive selection reduced bias at no cost in
variance relative to the fixed-propensity weights.

In the Ontario data the temperature effect is large, positive, and highly
significant across all thirteen schemes, and the estimators that perform best in
simulation place it near $9.8$ MW per squared degree Celsius, close to the unadjusted
value of $9.34$. Single-lag confounding adjustment therefore changes the estimate
only modestly in these data, which indicates that the confounding of the quadratic
temperature effect by recent weather, though clearly present, is smaller at the
single-lag horizon than in the deliberately strong simulation design. The cumulative
three-day estimators return smaller values, but they rest on roughly one-sixth of the
effective sample size of the single-lag schemes and are the schemes most exposed to
the air-density overlap problem, so their attenuation reflects a positivity
limitation rather than a more complete removal of confounding. The strengths of the
analysis follow from this design: the simulation is calibrated to the data and
reports Monte Carlo standard errors, so that estimator comparisons are interpretable
against simulation noise \citep{morris2019using}, and inference respects the
single-series structure through HAC and block-bootstrap variance estimation rather
than independence-based errors, with the adaptive families receiving bootstrap
intervals that account for the selection step.

The principal limitation is a positivity constraint that is structural rather than
statistical, since air density is by the ideal-gas relation nearly a deterministic
function of temperature, so that conditioning on lagged air density leaves little
independent variation in the exposure. No weighting scheme removes its imbalance, and
because the cumulative schemes multiply three treatment models they are the most
affected, a manifestation of the well-documented instability of inverse-probability
estimators under weak overlap \citep{kang2007demystifying, petersen2012diagnosing}.
This limitation, together with the two-year span that constrains precision and
precludes the study of seasonal effect modification, defines the agenda for
extending the analysis. Flexible, machine-learning estimation of the continuous
propensity density could better capture the nonlinear dependence among lagged weather
variables while retaining valid weights \citep{lee2010improving, kennedy2017nonparametric},
and the near-positivity induced by air density argues both for excluding air density
on physical grounds, since a deterministic function of temperature cannot satisfy
overlap, and for estimands and estimators designed to be robust to weak overlap, such
as overlap weights and trimmed targets \citep{li2018balancing, crump2009dealing}; a
longer series would in addition allow the joint three-day regime to be studied with
adequate overlap. Methodologically, the results show that outcome-adaptive and
fused-LASSO confounder selection transfers from binary to continuous longitudinal
treatments while preserving the efficiency rationale of the original method, and that
stabilized weighting remains the most reliable choice under near-positivity.
Substantively, the causal temperature effect is close to but not identical with the
unadjusted association, so demand models that ignore weather confounding will be
modestly biased for the quantity relevant to scenario planning, and the continuous
LOAL and fused-LOAL estimators developed here provide a route to correcting that bias
in energy-demand analysis.

\section*{Acronyms}
\label{sec:acronyms}

\begin{table}[htbp]
\centering
\renewcommand{\arraystretch}{1.15}
\begin{tabular}{ll}
\toprule
\textbf{Acronym} & \textbf{Definition} \\
\midrule
AR(1)  & first-order autoregressive process \\
ARIMA  & autoregressive integrated moving average \\
BIC    & Bayesian information criterion \\
DAG    & directed acyclic graph \\
FLOAL  & fused longitudinal outcome-adaptive LASSO \\
HAC    & heteroskedasticity- and autocorrelation-consistent \\
IPW    & inverse-probability weighting \\
LASSO  & least absolute shrinkage and selection operator \\
LOAL   & longitudinal outcome-adaptive LASSO \\
MCSD   & Monte Carlo standard deviation \\
MSE    & mean squared error \\
MSM    & marginal structural model \\
OAL    & outcome-adaptive LASSO \\
sIPW   & stabilized inverse-probability weighting \\
\bottomrule
\end{tabular}
\end{table}

\section*{Acknowledgements}
\label{sec:ack}

This work was supported by Dr.\ Kalia's research grant and by funding from the
Department of Statistics and the Faculty of Science at the University of Manitoba.

\section*{Data Availability, Software, and Code}
\label{sec:data}

All analyses were conducted in \textsf{R}. The simulation and application code,
including the data-generating process, the continuous-treatment LOAL and fused-LOAL
routines, the g-computation of the true value, and the HAC and moving-block bootstrap
inference, is provided in the accompanying script \texttt{R-script.Rmd} and was used
to produce Tables~\ref{tab:sim_all} and~\ref{tab:msm_results} and
\cref{fig:Balance assessment}. The Ontario weather series is available from the
Renewables.ninja reanalysis service (\url{https://www.renewables.ninja}) and the
demand series from the Independent Electricity System Operator
(\url{https://www.ieso.ca}).

\bibliographystyle{apalike}
\bibliography{references}

\newpage
\begin{appendices}
\section{Appendix}
\label{app:appendix}

Section~\ref{app:notation} fixes notation,
Section~\ref{app:assumptions} states the identification and estimation conditions,
Section~\ref{app:identification} derives identification of the marginal structural
parameter for a continuous treatment and gives the single-series justification of
the estimator, Section~\ref{app:weights} establishes the mean-one property of the
stabilized weights, and Section~\ref{app:dgp} specifies the simulation
data-generating process.

\subsection{Notation}
\label{app:notation}

Days are indexed $t = 1,\dots,T$. For a time-indexed variable $Z$, $Z_t$ is its
value on day $t$ and $\bar Z_t = (Z_1,\dots,Z_t)$ its history through day $t$. The
treatment $A_t \in \mathcal A \subseteq \mathbb R$ is daily mean temperature, the
outcome $Y_t$ is daily mean demand, $\boldsymbol L_t \in \mathbb R^{d}$ collects the
time-varying weather confounders, and $\boldsymbol V_t = (D_t, S_t)$ collects the
exogenous predictors that are not descendants of treatment. The observed data on day
$t$ are $\mathcal{O}_t = \{A_t, \boldsymbol L_t, Y_t, D_t, S_t\}$, and the analysis
uses the single realized path $\bar{\mathcal{O}}_T = (\mathcal{O}_1,\dots,
\mathcal{O}_T)$. A candidate regime is $\bar a = (a_1,\dots,a_t)$, and $Y_t(\bar a)$
is the potential demand on day $t$ under regime $\bar a$
\citep{bojinov2019time, robins1986new}. We write $f^{\mathcal O}$ for a density under
the observed (observational) law and $f^{\mathcal E}$ for a density under a
hypothetical randomized law in which treatment is assigned independently of
$\boldsymbol L$, with $\E^{\mathcal O}$ and $\E^{\mathcal E}$ the corresponding
expectations. For a continuous treatment $f(a_t\mid\cdot)$ is a conditional density,
$\one\{\cdot\}$ an indicator, and $\ind$ denotes conditional independence. To keep
the derivation transparent we present the three-time-point case $k = 0,1,2$; the
general-$T$ result follows by the identical factorization over $t = 1,\dots,T$.

\subsection{Identification and estimation conditions}
\label{app:assumptions}

Identification uses the standard assumptions stated in the main text
(Section~\ref{sec:assumptions}), restated here for the continuous, longitudinal
setting. Consistency requires that $\bar A_t = \bar a$ imply $Y_t = Y_t(\bar a)$,
with $Y_t$ depending on the regime only through the treatment path. Sequential
exchangeability requires $Y_t(\bar a) \ind A_s \mid \bar A_{s-1},
\bar{\boldsymbol L}_s$ for every $s \le t$ and every regime $\bar a$, and positivity
requires $f_{A_s\mid \bar A_{s-1},\bar{\boldsymbol L}_s}(a_s) \ge \varepsilon > 0$ for
every $s$ and every $a_s$ in the support of interest, which guarantees that the
density-ratio weights below are well defined and have finite variance
\citep{robins2000marginal, cole2008constructing}. Estimation from a single realized
path additionally requires that the joint process
$\{(Y_t,A_t,\boldsymbol L_t,\boldsymbol V_t)\}_{t\ge1}$ be strictly stationary and
$\alpha$-mixing, with mixing coefficients decaying fast enough that a law of large
numbers and a central limit theorem hold for sample averages of the weighted
estimating functions; this condition is a property of the data process rather than a
causal assumption, and it replaces the independent-replicates argument of the panel
setting \citep{newey1987simple, bojinov2019time, blackwell2018tscs}.

\subsection{Identification of the marginal structural parameter}
\label{app:identification}

\paragraph{Factorization of the two laws.}
Under the observational law, the joint density of
$(\bar{\boldsymbol L}_2,\bar A_2,\bar Y_2)$ factorizes as
\begin{align}
f^{\mathcal{O}} &= f^{\mathcal{O}}_{L_0}(l_0)\,
  f^{\mathcal{O}}_{A_0 \mid L_0}(a_0 \mid l_0)\,
  f^{\mathcal{O}}_{Y_1 \mid L_0,A_0}(y_1 \mid l_0,a_0)\,
  f^{\mathcal{O}}_{L_1 \mid L_0,A_0}(l_1 \mid l_0,a_0) \notag\\
  &\quad\times f^{\mathcal{O}}_{A_1 \mid L_0,L_1}(a_1 \mid l_0,l_1)\,
  f^{\mathcal{O}}_{Y_2 \mid L_1,A_1,Y_1}(y_2 \mid l_1,a_1,y_1)\,
  f^{\mathcal{O}}_{L_2 \mid L_1,A_1}(l_2 \mid l_1,a_1) \notag\\
  &\quad\times f^{\mathcal{O}}_{A_2 \mid L_0,L_1,L_2}(a_2 \mid l_0,l_1,l_2)\,
  f^{\mathcal{O}}_{Y_3 \mid L_2,A_2,Y_1,Y_2}(y_3 \mid l_2,a_2,y_1,y_2).
\end{align}
The hypothetical randomized law $f^{\mathcal{E}}$ is identical except that each
conditional treatment density $f^{\mathcal{O}}_{A_k\mid \bar{\boldsymbol L}_k}$ is
replaced by an assignment density $g_{A_k}(a_k)$ that does not depend on
$\boldsymbol L$; in the stabilized construction $g_{A_k}$ is the marginal treatment
density $f^{\mathcal{O}}_{A_k\mid\bar A_{k-1}}$,
\begin{align}
f^{\mathcal{E}} &= f^{\mathcal{O}}_{L_0}(l_0)\,
  g_{A_0}(a_0)\,
  f^{\mathcal{O}}_{Y_1 \mid L_0,A_0}(y_1\mid l_0,a_0)\,
  f^{\mathcal{O}}_{L_1 \mid L_0,A_0}(l_1\mid l_0,a_0)\notag\\
  &\quad\times g_{A_1}(a_1)\,
  f^{\mathcal{O}}_{Y_2\mid L_1,A_1,Y_1}(y_2\mid l_1,a_1,y_1)\,
  f^{\mathcal{O}}_{L_2\mid L_1,A_1}(l_2\mid l_1,a_1)\notag\\
  &\quad\times g_{A_2}(a_2)\,
  f^{\mathcal{O}}_{Y_3\mid L_2,A_2,Y_1,Y_2}(y_3\mid l_2,a_2,y_1,y_2).
\end{align}
Because all non-treatment factors are shared, the Radon-Nikodym derivative of the
randomized law with respect to the observational law is a product of treatment
density ratios and does not involve a measure-zero event,
\begin{equation}
\label{app:eq-rn}
\frac{\d f^{\mathcal{E}}}{\d f^{\mathcal{O}}}
   \;=\; \prod_{k=0}^{2}
   \frac{g_{A_k}(a_k)}
        {f^{\mathcal{O}}_{A_k\mid \bar{\boldsymbol L}_k}(a_k\mid \bar l_k)}
   \;=:\; W ,
\qquad
\bar{\boldsymbol L}_0 = l_0,\ \bar{\boldsymbol L}_1=(l_0,l_1),\
\bar{\boldsymbol L}_2=(l_0,l_1,l_2).
\end{equation}
Setting $g_{A_k}\equiv f^{\mathcal O}_{A_k\mid\bar A_{k-1}}$ gives the stabilized
weight $\mathrm{SW}$ of \eqref{eq:weights}, and setting $g_{A_k}$ to a constant gives
the unstabilized weight up to a normalizing constant that cancels in the Hájek
estimator below. The density ratio \eqref{app:eq-rn} replaces the indicator
$\one\{\bar A = \bar a\}$ used for a discrete treatment, so the argument applies to a
continuous treatment, for which $\PP(\bar A = \bar a)=0$.

\paragraph{Identification of the MSM.}
For any integrable function $h$, importance reweighting gives
$\E^{\mathcal E}[h] = \E^{\mathcal O}[\,W\,h\,]$. Under consistency, sequential
exchangeability, and positivity, treatment is independent of $\boldsymbol L$ under
$f^{\mathcal E}$, so the reweighted pseudo-population association between $Y_t$ and
the exposure equals the causal dose-response and the marginal structural model
\eqref{eq:msm} holds in that pseudo-population
\citep{robins2000marginal, hernan2001marginal}. Its parameters therefore solve the
weighted population estimating equation
\begin{equation}
\label{app:eq-ee}
\E^{\mathcal{O}}\!\Big[\,W_t\,\big\{Y_t - m(\boldsymbol X_t;\boldsymbol\beta)\big\}\,
   \boldsymbol X_t\Big] = \boldsymbol 0,
\qquad
m(\boldsymbol X_t;\boldsymbol\beta)=\beta_0+\beta_1(a_{t-1}-c)^2+\beta_2 D_t+\beta_3 S_t,
\end{equation}
with $\boldsymbol X_t = (1,(a_{t-1}-c)^2,D_t,S_t)^{\!\top}$ and $W_t$ the single-lag
weight \eqref{eq:weights}, which defines $\beta_1$ as the target marginal structural
dose-response parameter.

\paragraph{Estimation from a single series.}
Replacing the population expectation in \eqref{app:eq-ee} by the sample average over
the one observed path yields the weighted least-squares estimator. The population
mean under a fixed reference regime, used to define the true value in the simulation,
is the Hájek self-normalized functional
\begin{equation}
\label{app:eq-hajek}
\mu(\bar a) \;=\; \frac{\E^{\mathcal O}[\,W\,Y_3\,]}{\E^{\mathcal O}[\,W\,]},
\qquad
\hat\mu(\bar a) \;=\; \frac{\sum_{t} W_t\,Y_t}{\sum_{t} W_t},
\end{equation}
where the normalization by $\sum_t W_t$ is retained throughout, correcting the
Horvitz-Thompson form $n^{-1}\sum_t W_t Y_t$, which is valid only when
$\E^{\mathcal O}[W]=1$ exactly. Consistency of $\hat\mu(\bar a)$ and of the weighted
least-squares solution to \eqref{app:eq-ee} follows not from a law of large numbers
over independent subjects, since there is only one unit, but from the ergodic theorem
applied to the strictly stationary, $\alpha$-mixing process, under which time
averages converge to their expectations \citep{bojinov2019time, newey1987simple}. The
corresponding central limit theorem for dependent data justifies the HAC variance
estimator used for the fixed-propensity family and the moving-block bootstrap used
for the selection-based family \citep{newey1987simple, schnitzer2026adaptive}.

\subsection{The weights and their mean-one property}
\label{app:weights}

For the stabilized weight, iterated expectations give, at each lag $k$,
\begin{align}
\E^{\mathcal O}\!\left[
  \frac{f^{\mathcal O}_{A_k\mid\bar A_{k-1}}(A_k)}
       {f^{\mathcal O}_{A_k\mid\bar A_{k-1},\bar{\boldsymbol L}_k}(A_k)}
  \;\Bigg|\; \bar A_{k-1},\bar{\boldsymbol L}_k\right]
= \int
  \frac{f^{\mathcal O}_{A_k\mid\bar A_{k-1}}(a)}
       {f^{\mathcal O}_{A_k\mid\bar A_{k-1},\bar{\boldsymbol L}_k}(a)}\,
  f^{\mathcal O}_{A_k\mid\bar A_{k-1},\bar{\boldsymbol L}_k}(a)\,\d a
= 1 ,
\end{align}
so that, by the tower property applied sequentially over $k=0,1,2$,
$\E^{\mathcal O}[\mathrm{SW}] = 1$. This mean-one property, which underlies the Hájek
normalization and the variance reduction of $\mathrm{SW}$ relative to the
unstabilized weight, requires stationarity: under non-stationarity the conditional
treatment densities are not constant over time, the equality can fail, and estimates
from a non-stationary series can be spurious \citep{box1976analysis, hamilton2020time},
which is why stationarity is assessed empirically before estimation.

\subsection{Data-generating process for the simulation}
\label{app:dgp}

The simulation of Section~\ref{sec:sim} is calibrated to the Ontario series and,
unlike a design in which treatment is generated independently of the confounders,
induces confounding together with treatment-confounder feedback. For $t=1,\dots,N$
with $N = 730$, let $s_t = \cos\!\big(2\pi(d_t-172)/365.25\big)$ denote the seasonal
signal and $D_t$ the day type. Four representative time-varying confounders
$\boldsymbol L_t=(L_{1t},\dots,L_{4t})$ evolve as autoregressive processes with a
seasonal mean and feedback from the previous day's temperature,
\begin{equation}
\label{app:eq-L}
L_{jt} \;=\; \nu_j + \phi_j\,L_{j,t-1} + \eta_j\,s^{(j)}_t
             + \kappa_j\,A_{t-1} + \xi_{jt},
\qquad \xi_{jt}\sim\mathcal N(0,\sigma_{L_j}^2),
\end{equation}
with the binary confounders obtained by thresholding the corresponding latent
process. Temperature depends on the contemporaneous confounders, the seasonal signal,
and noise,
\begin{equation}
\label{app:eq-A}
A_t \;=\; a_0 + \delta\,s_t + \sum_{j=1}^{4}\theta_j\,L_{jt} + \varepsilon^{A}_t,
\qquad \varepsilon^{A}_t\sim\mathcal N(0,\sigma_A^2),
\end{equation}
and demand depends on the previous day's temperature through a centered quadratic, on
the lagged confounders, and on its own recent lags,
\begin{equation}
\label{app:eq-Y}
Y_t \;=\; b(D_t) + \beta_1\,(A_{t-1}-c)^2 + \sum_{j=1}^{4}\psi_j\,L_{j,t-1}
          + \rho_1 Y_{t-1} + \rho_2 Y_{t-2} + \varepsilon^{Y}_t,
\qquad \varepsilon^{Y}_t\sim\mathcal N(0,\sigma_Y^2).
\end{equation}
Because $\boldsymbol L_{t-1}$ enters both \eqref{app:eq-A}, through $A_{t-1}$, and
\eqref{app:eq-Y}, it confounds the $A_{t-1}\to Y_t$ effect, the loadings $\theta_j$
creating the $\boldsymbol L\to A$ arrows, and because $A_{t-1}$ enters
\eqref{app:eq-L} the confounders are mediators of earlier temperature through the
feedback $\kappa_j$, reproducing the structure of \cref{fig:DAG_EO}. The treatment
noise variance $\sigma_A^2$ is set large enough that positivity holds by
construction, so the simulation isolates estimator performance under valid overlap,
whereas the near-positivity of the application arises from the near-deterministic
dependence of air density on temperature that the simulation does not impose. The
calibrated values are listed in Table~\ref{tab:params}. The true value of $\beta_1$
is obtained by g-computation, intervening on $A_t$ by severing \eqref{app:eq-A},
propagating \eqref{app:eq-L} to \eqref{app:eq-Y}, and averaging the potential
outcomes over a temperature grid; under the additive specification it coincides with
the structural coefficient $\beta_1$, which we verify numerically.

\begin{table}[ht]
\centering
\caption{Data-generating process for the simulation, calibrated to Ontario. Effect
sizes $\beta_1\in\{0,2,4\}$, and the centering constant $c$ is the demand-minimizing
temperature. The confounder-to-treatment loadings $\theta_j$ and the feedback
loadings $\kappa_j$ create the confounding and the treatment-confounder feedback,
respectively.}
\label{tab:params}
\renewcommand{\arraystretch}{1.3}
\small
\begin{tabular}{>{\RaggedRight}p{2.4cm} >{\RaggedRight}p{2.6cm}
 >{\RaggedRight}p{8.5cm}}
\toprule
\textbf{Component} & \textbf{Symbol} & \textbf{Specification} \\
\midrule
Temperature (treatment) & $A_t$, Eq.~\eqref{app:eq-A} &
Baseline $a_0 = 16$; seasonal loading $\delta = 10$; confounder loadings
$\theta = (-0.8,\,-15,\,-3,\,-2)$ on contemporaneous $L_{jt}$; residual s.d.\
$\sigma_A = 9$. \\
Demand (outcome) & $Y_t$, Eq.~\eqref{app:eq-Y} &
Day-type baseline $b(D_t) = 45{,}000$ (weekday and weekend), $43{,}000$ (holiday);
causal curvature $\beta_1 \in \{0,2,4\}$; lagged confounder loadings
$\psi = (-3000,\,-4500,\,-12000,\,-8000)$; autoregression $\rho_1 = 0.10$,
$\rho_2 = 0.02$; residual s.d.\ $\sigma_Y = 200$. \\
Confounder 1 (precipitation) & $L_{1t}$, Eq.~\eqref{app:eq-L} &
AR(1) $\phi_1 = 0.50$; base $\nu_1 = 0.8$; seasonal loading on $s_t$; feedback
$\kappa_1 = 0.02$; residual s.d.\ $0.3$; truncated at zero. \\
Confounder 2 (air density) & $L_{2t}$ &
AR(1) $\phi_2 = 0.45$; base $\nu_2 = 0.69$; seasonal loading on $s_t$; feedback
$\kappa_2 = 0.001$; residual s.d.\ $0.02$. \\
Confounder 3 (snowfall, binary) & $L_{3t}$ &
Latent AR(1) $\phi_3 = 0.70$ with seasonal step and melt term; feedback
$\kappa_3 = 0.28$; residual s.d.\ $10$; thresholded at $12$. \\
Confounder 4 (snow mass, binary) & $L_{4t}$ &
Latent AR(1) $\phi_4 = 0.60$; seasonal loading on $s_t$; feedback $\kappa_4 = 0.25$;
residual s.d.\ $15$; thresholded at $75$. \\
Day type & $D_t$ &
Weekday, weekend, or statutory holiday (Ontario, 2018 to 2019). \\
\bottomrule
\end{tabular}
\end{table}

The causal structure assumed for the application is summarized in \cref{fig:DAG_1},
a conceptual diagram of the roles of the variables rather than a fully time-expanded
graph, for which \cref{fig:DAG_EO} is the reference.

\begin{landscape}
\begin{figure}
    \centering
\scalebox{0.85}{
\begin{tikzpicture}[
                > = {Latex[length=2mm, width=2mm]}, 
                shorten > = 1pt, 
                auto,
                scale = 1.2,
                node distance = 3cm, 
                semithick 
            ]
    
            \tikzstyle{state}=[
                rectangle,
                draw = black,
                rounded corners = .25cm,
                thick,
                minimum size = 9.5mm,
                inner sep = 8pt,
                align = center
            ]
            
            \tikzstyle{confounder}=[
                rectangle,
                draw = black,
                rounded corners = .25cm,
                thick,
                minimum size = 9.5mm,
                inner sep = 8pt,
                fill = gray!30,
                align = center
            ]
    \node[confounder] (weather) at (0, 4) {
        \textcolor{black}{
        \begin{tabular}{c}
       Past Weather Variables$^a$ \\
        (Precipitation, Snowfall, \\
        Snow Mass, Cloud Cover, \\
        Air Density)
        \end{tabular}
        }
    };
    
    \node[state] (temperature) at (-3, 0) {
        \textcolor{black}{
        \begin{tabular}{c}
        Temperature \\
        (Treatment)
        \end{tabular}
        }
    };
    
    \node[state] (demand) at (3, 0) {
        \textcolor{black}{
        \begin{tabular}{c}
        Electricity Demand \\
        (Outcome)
        \end{tabular}
        }
    };
    
    \node[state] (daytype) at (0, -3) {
        \textcolor{black}{
        \begin{tabular}{c}
        Day Type \\
        (Weekday/Weekend),\\
        Current Weather Variables \\
        (Precipitation, Snowfall, \\
        Snow Mass, Cloud Cover, \\
        Air Density)
        \end{tabular}
        }
    };
    \draw[->, thick] (weather) -- (-1.5, 2) -- (temperature);
    \draw[->, thick] (weather) -- (1.5, 2) -- (demand);
    \draw[->, thick] (temperature) -- (demand);
    \draw[->, thick] (daytype) -- (demand);
                
\end{tikzpicture}
}
\caption{Conceptual causal structure for the electricity-demand application:
temperature (treatment), past weather variables (time-varying confounders, taken as
lagged values at one to three days), and current weather and day type (additional
outcome predictors that are not confounders). Arrows indicate the direction of causal
influence. This is a schematic summary rather than a fully time-expanded directed
acyclic graph.}
\label{fig:DAG_1}
\end{figure}
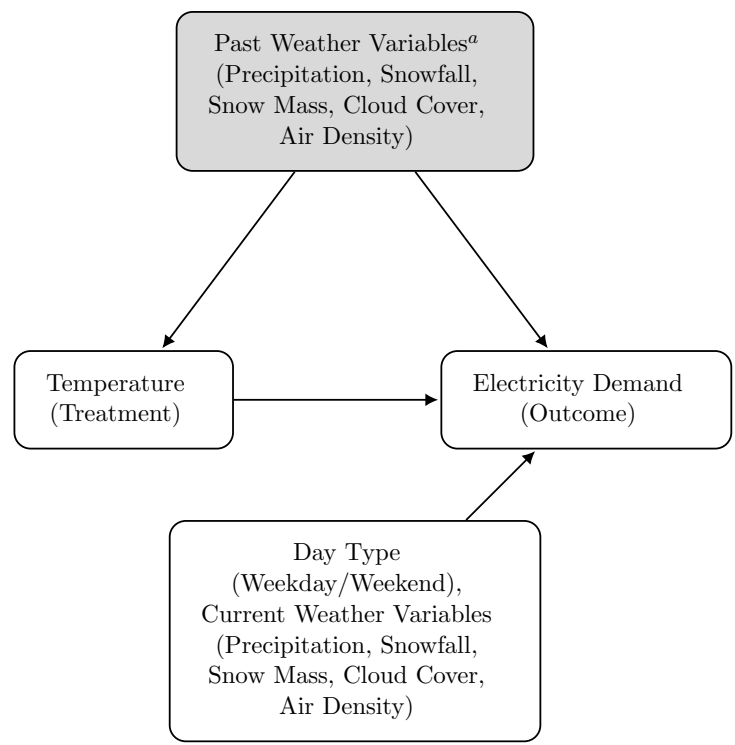
\end{landscape}

\begin{figure}[htbp]
\centering
\includegraphics[width=\textwidth]{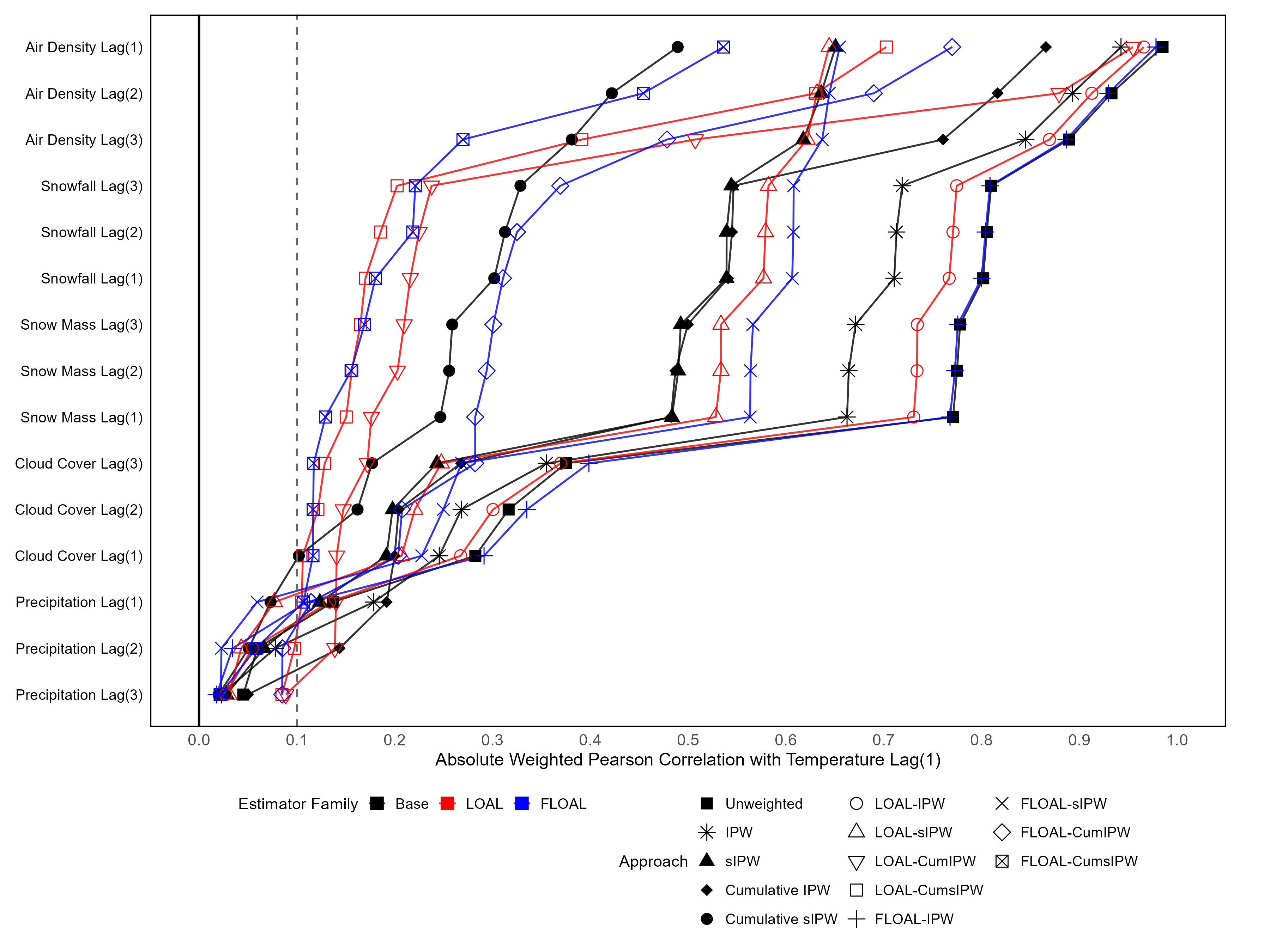}
\caption{Covariate balance for the Ontario application under the thirteen weighting
schemes. Each point is the absolute weighted correlation between the exposure
$A_{t-1}$ and a lagged confounder (air density, snowfall, snow mass, cloud cover, and
precipitation at lags 1--3); covariates are ordered by their unweighted correlation.
The dashed line marks the balance threshold $|r| = 0.1$.}
\label{fig:Balance assessment}
\end{figure}
\end{appendices}

\end{document}